\documentclass[graybox]{svmult}

\usepackage{mathptmx}      
\usepackage{helvet}            
\usepackage{courier}           
\usepackage{type1cm}        
\usepackage{makeidx}        
\usepackage{graphicx}        
\usepackage{multicol}         
\usepackage[bottom]{footmisc}
\usepackage{amssymb}

\def\Pa       		{{\bf P}_{a}}

\def\GA       		{{\bf G}_{A}}
\def\Jab	    	{{\bf J}_{ab}}
\def\Zab			{{\bf Z}_{ab}}

\def\Zabcde			{{\bf Z}_{abcde}}
\def\Palbe			{{\bf P}_{\alpha\beta}}
\def\Qal	    	{{\bf Q}_{\alpha}}
\def\Qbe			{{\bf Q}_{\beta}}
\def\eIa	    	{{\epsilon_I}^\alpha}
\def\eIb		    {{\epsilon_I}^\beta}
\def\eJa		    {{\epsilon_J}^\alpha}
\def\eJb		    {{\epsilon_J}^\beta}
\def\uIa		    {{u^I}_\alpha}
\def\uIb		    {{u^I}_\beta}
\def\uJa		    {{u^J}_\alpha}

\def\lal		    {{\lambda}_\alpha}
\def\lbe		    {{\lambda}_\beta}
\def\wal		    {w^\alpha}
\def\wbe	    	{w^\beta}
\def\preon	    	{|\lambda\rangle}
\def\covD	    	{\mathcal{D}}
\def\Ralal	    	{{\mathcal{R}_\alpha}^\alpha}
\def\Ralbe	    	{{\mathcal{R}_\alpha}^\beta}
\def\Ralga	    	{{\mathcal{R}_\alpha}^\gamma}
\def\Rgabe    		{{\mathcal{R}_\gamma}^\beta}
\def\Oalal	    	{{\Omega_\alpha}^\alpha}
\def\Oalbe	    	{{\Omega_\alpha}^\beta}
\def\Oupalbe		{\Omega^{\alpha\beta}}
\def\Oupbeal		{\Omega^{\beta\alpha}}
\def\Oskalbe		{\Omega^{[\alpha\beta]}}
\def\Oalga			{{\Omega_\alpha}^\gamma}
\def\Ogabe			{{\Omega_\gamma}^\beta}
\def\Oosp			{\Omega^{osp(32|1)}}
\def\Ocjs			{\Omega^{CJS}}
\def\Omth			{\Omega^{M-algebra}}
\def\Ga				{{\Gamma^a}}
\def\Gab			{{\Gamma^{ab}}}
\def\Gbbb			{{\Gamma^{{b_1}{b_2}{b_3}}}}
\def\Gabbbb			{{\Gamma^{a{b_1}{b_2}{b_3}{b_4}}}}
\def\Gabcde			{{\Gamma^{abcde}}}
\def\Fabbb			{F_{a{b_1}{b_2}{b_3}}}
\def\Fbbbb			{F_{{b_1}{b_2}{b_3}{b_4}}}

\makeindex     

\begin{document}

\title*{Sources for Chern-Simons theories}
\author{Jos\'e D. Edelstein and Jorge Zanelli}
\institute{Jos\'e D. Edelstein \at Departament of Particle Physics and IGFAE,
University of Santiago de Compostela, Spain \\
Centro de Estudios Cient\'{\i}ficos, CECS, Valdivia, Chile \\
\email{jedels-at-usc-dot-es}
\and Jorge Zanelli \at Centro de Estudios Cient\'{\i}ficos, CECS,
Valdivia, Chile \\ \email{z-at-cecs-dot-cl}}
%
%
\maketitle

\abstract{The coupling between Chern-Simons theories and matter sources defined by branes of different dimensionalities is examined. It is shown that the standard coupling to membranes, such as the one found in supergravity or in string theory, does not operate in the same way for CS theories; the only $p$-branes that naturally couple seem to be those with $p=2n$; these $p$-branes break the gauge symmetry (and supersymmetry) in a controlled and sensible manner.}

\section{Introduction} 
\label{Introduction}

Chern-Simons (\textbf{CS}) theories have a number of appealing attributes that make them interesting candidates for the description of natural phenomena. In spite of their promise, they also present a number of puzzling features that set them in a different class from other gauge theories, such as those that have been successfully used for the description of the Standard Model. In the case of higher (than three) dimensional CS theories, aiming at describing gravitational physics, several important difficulties emerge. Prominent among these, stands the problem of how to couple them to different forms of matter such as, for instance, branes of different dimensionalities.

These extended objects, whose existence is familiar in the context of string theory, play a natural role in CS gravitational theories based on supersymmetric extensions of both the Poincar\'e and the anti de Sitter (\textbf{AdS}) groups. In the particular case of 11D, while M-theory is well-known to possess two kinds of branes --the electric M2- and the magnetic M5-branes--, the very same objects naively appear in a CS theory based on the M-algebra \cite{HTZ}. It is not clear, however, whether these objects actually belong to the spectrum of CS theory, and how they couple to the remaining fields. We shall address both problems, on general grounds, in the present article.

Interactions with sources provide a handle to probe the perturbative structure of quantum theories, but that requires a well defined expansion of the interacting theory as a power series in a weak coupling parameter. In this manner, currents generated by point charges --and more generally, by charged extended objects such as strings or higher dimensional branes--, are standard mechanisms that allow extracting predictions from the effective low energy limit in string/supergravity theories, which are in principle experimentally testable. 

On the other hand, the perturbative expansion as a power series in the coupling constant seems to be of little use in a CS system: CS theories are highly nonlinear and self-interacting in a way that it is not possible to distinguish between the ``free'' and the ``interacting'' parts of the action without making a severe mutilation of the system.  CS Lagrangians have no adjustable coupling constants (dimensionful or otherwise). This a priori appealing feature has a downside: the separation between background and perturbations is not clear-cut either. 

A further complication is that since CS actions do not involve a metric, there is no notion of energy, and hence no energy scale is naturally defined in them. An energy scale can be introduced only at the cost of breaking gauge symmetry. In this sense, CS systems can be viewed as the analog of noble gases in chemistry, because they would not interact or bind to any other form of matter. It may seem as if they are inert, subtle beautiful structures to be admired, unrelated to the physical reality of the world. Here we will argue that this is not quite true: there is no obstruction to the coupling between membranes whose worldvolumes are odd-dimensional and non-Abelian CS systems.

On the road to understand such couplings, one faces the issue of uncovering the BPS spectrum of supersymmetric higher dimensional CS theories. Besides the expected presence of BPS branes preserving one half or one quarter of the original supersymmetries, it is interesting to seek for possible states preserving all but one supercharges that may be understood as constituents of the former. These so-called BPS preon states were proposed in \cite{BdAIL}, and they were recently alleged to exist in the $osp(32|1)$ CS theory (and presumably in other related theories) \cite{BdAIPV}.  The existence of preons has also been recently ruled out as solutions of the presumed low-energy limit of M-theory, the Cremmer-Julia-Scherk (\textbf{CJS}) supergravity \cite{GGPR}. Here we show that the arguments presented in \cite{BdAIL} are appropriate as well for CS theories based on extensions of the AdS algebra, but they need a subtle improvement for the case of Poincar\'e based theories such as the CS supergravity for the M-algebra.

\vskip3mm
-----------------------------\\[-.7em]

Claudio Bunster was a promoter of the idea that the world history of a point particle, as well as that of the entire Universe, can be viewed as similar objects, to be treated quantum mechanically in a similar way \cite{CT82}. Bunster was also a pioneer in considering currents with support on branes as sources coupled to $p$-form gauge potentials. In \cite{Teitelboim:1985ya}, he showed that it is impossible to minimally couple a non-Abelian connection to a $p$-brane for $p>0$. It is therefore a suitable form of tribute to celebrate his sixtieth birthday, to discuss a context in which this obstruction can be circumvented.

\section{Remarks on the BPS spectrum of CS supergravity} 
\label{Remarks}

A key feature of higher dimensional supergravity theories (and, more generally, of string/M-theory) is their natural coupling to certain branes. The dimensionality of these objects is strongly restricted by the tensorial properties of the field content of the theory. In 11D supergravity, for instance, the three-form field couples naturally to an electric 2-brane (M2) or to a magnetic 5-brane (M5). Moreover, these branes are BPS states, their mass being quantum mechanically equal to their charges.

The BPS spectrum of higher dimensional CS theories is not yet well understood. There are some scattered examples in the literature but no general results or exhaustive studies have been undertaken so far. In the present section we will illustrate some of the difficulties that this problem embodies. We will focus on the case of the CS supergravity for the M-algebra, though our results are quite general. We will argue that it may be necessary to reconsider the way sources and couplings come into place in these theories. A concrete proposal is then presented in the forthcoming section.

\subsection{CS supergravity for the M-algebra} 
\label{CSMalg}

One of the nicest features of CS supergravities is that, being gauge theories, their dynamical variables are connections living in a given Lie algebra. Their fiber bundle structure seems an auspicious starting point towards a quantization program. However, this is an intricate problem, mostly due to the existence of highly nontrivial vacua with radically different dynamical content and the lack of a perturbative expansion around many of them \cite{LecturesZ,EZsort}.

It is possible to write down a CS supergravity theory with the symmetry dictated by the so-called M-algebra explicitly realized off-shell \cite{HTZ}. Soon after the discovery of M-theory \cite{MTheory}, it was suggested that CS supergravity might provide a covariant non-perturbative formulation of quantum M-theory \cite{Troncoso:1997va} based upon $osp(32|1)$, the minimal supersymmetric extension of the AdS group in 11D. The observation that this theory violates parity conservation (a symmetry that, for consistency, should be present in M-theory \cite{WittenMthParity}), prompted the suggestion of a CS theory based on $osp(32|1) \times osp(32|1)$ \cite{Horava}. These CS theories have a number of nice features that include the presence of a central extension whose tensorial character matches that of an extended object like the M5-brane. The M2-brane, instead, enters the game in a less natural way. A CS theory based on the M-algebra puts both basic constituents of M-theory in a more democratic ground.
\vskip1mm

The M-algebra includes, apart from the Poincar\'e generators $\Jab$ and $\Pa$, a Majorana supercharge $\Qal$ and two additional bosonic generators, $\Zab$ and $\Zabcde$ that close the supersymmetry algebra \cite{PKT},
\begin{equation}
\begin{array}{ccl}
\{ \Qal,\Qbe \} &=& \left( C\Gamma^a \right)_{\alpha \beta}\, \Pa + (C\Gamma^{ab})_{\alpha\beta}\, \Zab + (C\Gamma^{abcde})_{\alpha\beta} \Zabcde \\[.7em]
&\equiv& \textbf{P}_{\alpha\beta}\;,
\end{array}
\label{MAlgebra}
\end{equation}
where the charge conjugation matrix $C$ is antisymmetric. The ``central charges'' $\Zab$ and $\Zabcde$ are tensors under Lorentz rotations but are otherwise Abelian generators. It must be stressed that the M-algebra is not the same as $osp(32|1)$, nor a subalgebra of the latter, and not even a contraction of it. The M-algebra can be obtained through an expansion of $osp(32|1)$, which corresponds to an analytic continuation of the Maurer-Cartan form \cite{HS,dAIPV}. This mechanism was used to obtain the actions for the corresponding algebras in \cite{EHTZ}.
\vskip1mm

The field content of the theory is thus given by a connection in the M-algebra,
\begin{equation}
\mathcal{A} = \frac12 \omega^{ab}\,\Jab + e^a\,\Pa + \frac{1}{\sqrt{2}}
\psi^\alpha\,\Qal + b_{[2]}^{ab}\,\Zab + b_{[5]}^{abcde}\,\Zabcde\;,
\label{Mconnection}
\end{equation}
where $e$ and $\omega$ describe the metric and affine features of the spacetime geometry (including torsion); $\psi$ is the gravitino, and $b_{[2]}$ and $b_{[5]}$ are Abelian gauge fields in antisymmetric tensor representations of the Lorentz group. The field strength $\mathcal{F} = d \mathcal{A} + \mathcal{A} \wedge \mathcal{A}$ reads
\begin{equation}
\mathcal{F} = \frac12 R^{ab}\,\Jab + \tilde T^a\,\Pa + \frac{1}{\sqrt{2}}
\covD\psi^\alpha\,\Qal + \tilde F^{ab}_{[2]}\,\Zab + \tilde
F^{abcde}_{[5]}\,\Zabcde\;,
\label{Mfieldstrength}
\end{equation}
where $R^{ab}=d\omega ^{ab}+\omega ^a_{~c} \wedge \omega ^{cb}$ is the curvature 2-form, and
\begin{equation}
\begin{array}{ccl}
& & \tilde T^a = \covD e^a - \displaystyle{\frac14} \bar\psi \Gamma^a \psi\;, \\ [1.1em]
& & \tilde F^{a_1 \cdots a_k}_{[k]} = \covD b_{[k]}^{a_1 \cdots a_k} - \displaystyle{\frac14} \bar\psi \Gamma^{a_1 \cdots a_k} \psi\;.
\end{array}
\end{equation}
It is important to specify at this point the expression for the covariant derivative acting on the gravitino, $\covD\psi^\alpha = d\psi^\alpha + \frac14 {\omega^\alpha}_\beta \psi^\beta$.
\vskip1mm

The CS form\footnote{In general, for a $2n-1$ theory, the Chern-Simons form is given as $d\mathcal{C}_{2n-1} = \langle \mathcal{F} \wedge \cdots \wedge \mathcal{F} \rangle$ ($n$ times), $\langle \cdots \rangle$ being an invariant tensor for the corresponding Lie algebra.} is the Lagrangian of the theory, constructed through the standard requirement that $d\mathcal{C}_{11} = \langle \mathcal{F} \wedge \cdots \wedge \mathcal{F} \rangle$, where the bracket $\langle \cdots \rangle$ stands for a multilinear form of the M-algebra generators $\GA$ whose only non-vanishing bosonic components are
\begin{equation}\label{invT}
\begin{array}{ccl}
& & \langle {\bf J}_{a_1 a_2}\, {\bf J}_{a_3 a_4}\, {\bf J}_{a_5 a_6}\,
{\bf J}_{a_7 a_8}\, {\bf J}_{a_9 a_{10}}\, {\bf P}_{a_{11}} \rangle
= \displaystyle{\frac{16}{3}}\,\epsilon_{a_1 a_2 a_3 a_4 a_5 a_6 a_7 a_8 a_9 a_{10} a_{11}} \;, \\ [1.1em]
& & \langle {\bf J}_{a_1 a_2}\, {\bf J}_{a_3 a_4}\, {\bf J}_{a_5 a_6}\,
{\bf J}_{a_7 a_8}\, {\bf J}^{a_9 a_{10}}\, {\bf Z}_{abcde} \rangle
= - \displaystyle{\frac{4\alpha}{9}}\,\epsilon_{a_1 a_2 a_3 a_4 a_5 a_6 a_7 a_{8} a b c}\,\delta^{a_9 a_{10}}_{d e} \;, \\ [1.1em]
& & \langle {\bf J}_{a_1 a_2}\, {\bf J}_{a_3 a_4}\, {\bf J}_{a_5 a_6}\,
{\bf J}^{a_7 a_8}\, {\bf J}^{a_9 a_{10}}\, {\bf Z}^{ab} \rangle = \displaystyle{\frac{16 (1 - \alpha)}{3}}\,\left[ \delta_{a_1 \dots a_6}^{a_7 \dots a_{10} a b} - \delta_{a_1 \dots a_4}^{a_9 a_{10} a b} \delta_{a_5 a_6}^{a_7 a_8}
\right].
\end{array}
\end{equation}
As a consequence of this, the equations of motion are quintic polynomials in the curvature,
\begin{equation}
\langle \mathcal{F} \wedge \cdots \wedge \mathcal{F}\; \GA \rangle = 0 \;.
\label{eomMalg}
\end{equation}
%
\subsection{Preons in the CS M-theory} 

The spectrum of BPS states in M-theory goes beyond the M2- and M5-brane. By algebraic reasoning, one would expect to have so-called preons. These are states that preserve the supersymmetric invariance due to all but one real supercharge component. For instance, in 11D, this amounts to $31$ real supercharges. It has been argued that CS supergravity possesses BPS preons in its spectrum \cite{BdAIPV}. That proof, however, is somewhat biased by the assumption that the relevant Lie algebra  is $osp(32|1)$. In this section we will closely follow the compelling algebraic reasoning presented in \cite{BdAIPV} in support of a proof of existence of BPS states in CS M-theory. 

Assume that we are interested in a BPS preon solution preserving $31$ real supercharges. This means that there are $31$ (real components of the generalized) Killing spinors, $\eJa$, $J = 1 \ldots 31$. They are defined in terms of the differential operator that generates the supersymmetry transformation of the gravitino, $\delta \psi^\alpha = \covD \eJa = d \eJa + \Oalbe \eJb$. There is a single bosonic spinor $\lal$, orthogonal to $\eJa$, $\eJa\,\lal = 0$, that characterizes the expected preonic state. It is natural, then, to call it $\preon$, which schematically satisfies $\eJa\,\Qal\,\preon = 0$, for all $J$. Thus, $\Qal\,\preon = \lal\,\preon$, and therefore, $\Palbe\,\preon = \lal\,\lbe\,\preon$. It is useful to complete the basis of spinors both with indices up, $\{\eJa,\wal\}$, and down, $\{\uJa,\lal\}$. It is always possible to choose both bases in such a way that they are orthogonal $\wal \lal = 1$, $\eJa \uIa = \delta_J^I$, $\wal \uJa = \eJa \lal = 0$.

The $\eJa$ are Killing spinors, $\covD\,\eJa = 0$, and since $d (\eJa\,\lal) = 0$, it turns out that $\covD\,\lal$ is proportional to $\lal$. Let us thus define the one-form $A$ as $\covD\,\lal = A\,\lal$. The application of two consecutive covariant derivatives yields $\covD\covD\,\lal = \Ralbe\,\lbe$, where $\Ralbe$ is the generalized curvature two-form
\begin{equation}
\Ralbe = d\,\Oalbe - \Oalga \wedge \Ogabe\;.
\label{Ralbe}
\end{equation}
On the other hand, by applying the exterior derivative to the remaining orthogonality relations, the expression for the covariant derivatives of the remaining spinors $\covD \uIa = B^I\, \lal$, where $B^I$ is a collection of $31$ 1-forms, and $\covD \wal = - A\, \wal - B^I\, \eIa$, are easily obtained. Performing now the same trick on $\wal \covD \lal$ and $\eJa \covD \lal$, one gets $\covD \covD \lal = d A\; \lal$. It is easy to play the same trick with $\uIa$, with the result $\covD \covD \uIa = \Ralbe\,\uIb = \nabla B^I \lal$, where $\nabla B^I = d B^I + A \wedge B^I$. All in all, we can decompose both $\Omega$ and $\mathcal{R}$ in the spinorial basis as \cite{BdAIPV}
\begin{equation}
\begin{array}{ccl}
& & \Oalbe = A\, \lal\,\wbe + B^I\, \lal\,\eIb - d \lal\,\wbe - d \uIa\,\eIb \;, \\ [1.1em]
& & \Ralbe = d A\, \lal\,\wbe + \nabla B^I\, \lal\,\eIb \;.
\end{array}
\end{equation}
Now, let us complete the argument used in \cite{BdAIPV} for the case of the smallest AdS superalgebra $osp(32|1)$. The bosonic part of the connection is a one-form in the subalgebra $sp(32)$
\begin{equation}
\Oosp = \frac{1}{2}\, e_a\,\Ga + \frac{1}{4}\,\omega_{ab}\,\Gab +
\frac{1}{240}\, b_{abcde}\,\Gabcde \;.
\end{equation}
Thus, $\Oupalbe = \Oupbeal$, that is, $\Oskalbe = 0$. Both $\Omega$ and $\mathcal{R}$ belong to $sp(32)$. Thus, they are traceless $\Oalal = \Ralal = 0$. This means $A = 0$, then $\Ralbe = d B^I \lal \eIb$. This implies that the generalized supercovariant curvature is nilpotent \cite{BdAIPV}
\begin{equation}
\Ralga \wedge \Rgabe = 0\;,
\label{nilpotent}
\end{equation}
due to the orthogonality $\eIa \lal = 0$. Now, since the gauge connection in the $osp(32|1)$ CS supergravity, $\mathcal{A}$ precisely matches $\Omega$, then $\mathcal{F} = \mathcal{R}$. The equation of motion (that formally looks like (\ref{eomMalg})) is, then, always satisfied for a BPS preonic configuration. Still, it is necessary to check the actual integrability of eq.(\ref{nilpotent}), as there might be, for instance, topological obstructions.

Now, coming back to the case of the M-algebra, it is important to stress that $\Omth = \frac{1}{4}\,\omega_{ab}\,\Gab$, which is not $\mathcal{A}$. Then, $\mathcal{R} = \frac12 R^{ab}\,\Jab \neq \mathcal{F}$. The nilpotency of $\mathcal{R}$, thus, does not guarantee a priori the solution of the CS equations of motion. This is a generic feature of all theories built as extensions of Poincar\'e CS supergravity (see, for example, \cite{MokRom}). Notice that this is qualitatively different to the behavior of $AdS$-based Lie algebras. The connection between both kinds of theories, though, is well understood \cite{EHTZ}. However, recalling that the nonvanishing components of the invariant tensor for all these theories look like those displayed earlier in (\ref{invT}), we can conclude that at least four factors in eq.(\ref{eomMalg}) admit the replacement $\mathcal{F} \to \mathcal{R}$, and this guarantees that the preonic configuration -- if no topological obstruction arises and it actually solves (\ref{nilpotent}) -- always satisfies the CS supergravity equations.

\subsection{Difficulties with the standard M-brane construction in CS theory}
\label{NoGo} 

In standard supergravity one typically ``deduces'' from a given SUSY algebra that there are BPS states. Applying the same analysis to a CS theory in a straightfoward manner meets with severe difficulties that cast doubt on the viability of the strategy. In order to fix ideas, let us begin by recalling how it is that standard supergravity couples to a membrane. The excercise will suggest why a similar strategy would not work for a CS supergravity. Let us present the argument through an example.

Consider a flat M2-brane extended in the $x^1$-$x^2$ plane. It should be associated with a non-zero value of $Z_{12}$ (the very presence of the M2-brane breaks the Lorentz group from $SO(1,10)$ to $SO(1,2) \times SO(8)$). Let us choose the Majorana representation in which $C = \Gamma^0$ [$(\Gamma^0)^2=1$]. In that case, for a static membrane,
\begin{equation}
\{ \Qal,\Qbe \} = \delta_{\alpha \beta} \, P_0 + (\Gamma^{012})_{\alpha \beta } Z_{12} \;.
\label{M2Algebra}
\end{equation}
In 11D, the Majorana spinors $\Qal$ are real. So, the left hand side is manifestly positive definite. The sign of $Z_{12}$ can be flipped by replacing a membrane by an anti-membrane. Instead, $P_0 \geq 0$. Thus, as a consequence of the positive definite bracket, using Witten-Olive's construction \cite{WO}, it turns out that
\begin{equation}
P_0 \geq |Z_{12}| \;.
\end{equation}
A BPS M2-brane is expected to saturate the bound, $P_0 = Z_{12}$,
\begin{equation}
\{ \Qal,\Qbe \} = P_{0} \left[ 1 \mp \Gamma^{012} \right]_{\alpha \beta} \;.
\label{BPSM2Algebra}
\end{equation}
Spinors $\epsilon$ satisfying $\Gamma^{012}\,\epsilon = \pm \epsilon$ are eigenspinors of $\{ \Qal,\Qbe \}$ with zero eigenvalue. These are the spinors corresponding to the $1/2$ unbroken supersymmetries. A similar argument holds for the M5-brane. This argument is {\it naively independent of the dynamics}, {\it i.e.} whether it is given by a CS Lagrangian or that of Cremmer-Julia-Scherk. We will come back to this point shortly.
\vskip1mm

In standard CJS supergravity, $\delta \psi = (d + \Ocjs)\, \epsilon$, with the connection given by the $32 \times 32$ matrix valued 1-form
\begin{equation}
\Ocjs = \frac{1}{4}\,\omega_{ab}\,\Gab + \frac{i}{18}\,e^a\,\Fabbb\,\Gbbb
+ \frac{i}{144}\,e_a\,\Fbbbb\,\Gabbbb \;.
\label{Ocjs}
\end{equation}
An M2-brane has non-vanishing $F_{012r}$, $r$ being the transverse radial direction. Imposing $\delta \psi = 0$ leads to three different equations, $\delta \psi_m = \delta \psi_r = \delta \psi_s = 0$, where $m = 0,1,2$, $r$ amounts for the radial direction and $s$ runs over transverse indices. The second equation just provides a differential equation that dictates the radial dependence $\epsilon(r)$. Now, since the spinor $\epsilon$ obeys a chirality prescription, we see that,
\begin{equation}
\begin{array}{ccl}
\delta \psi_m & = & \displaystyle{\frac{1}{4}}\, \omega_{m\;ab}\Gamma^{ab} \epsilon - \displaystyle{\frac{1}{12}}\, \epsilon_{mnp}\, \Gamma^{npr}\, F_{012r}\,\epsilon = 0 \;,\\ [1.1em]
\delta \psi_s & = & \displaystyle{\frac{1}{4}}\, \omega_{s\;ab}\Gamma^{ab} \epsilon + \displaystyle{\frac{1}{12}}\, \Gamma^{012r}_{~~~~~\;\;s}\, F_{012r}\,\epsilon = 0 \;,
\end{array}
\end{equation}
and we see that, provided the only non-vanishing components of the spin connection are $\omega_m^{mr}$ and $\omega_s^{sr}$, which is the case in standard supergravity for a natural D-brane ansatz, previous equations would only lead to non-trivial solutions provided, precisely, the chirality condition $\Gamma^{012}\,\epsilon = \pm \epsilon$ is imposed on the spinor. The whole picture is self-consistent.

Instead, in CS M-theory supergravity, the supersymmetry transformation of the gravitino is dramatically simpler, $\delta\psi_\mu = D_\mu \epsilon = 0$, the only difference having to do with the fact that now $\omega$ can have a contorsion contribution, $\omega = \omega^{(0)} + \kappa$. The naive expectation is that $\kappa$ should play the role of the $A_{[3]}$ form. For example, $A_{[3]} = e^a \wedge e^b \wedge \kappa_{ab}$ \cite{MaxB}. However, whatever is the case, the above equation would lead schematically to
\begin{equation}
\left( a_1\, \Gamma^{ab} + a_2\, \Gamma^{cd} \right) \epsilon = 0 \;,
\end{equation}
and this could never reduce to $\Gamma^{012}\,\epsilon = \pm \epsilon$, the projection that, according to our simple algebraic argument, is necessary for the M2-brane. At best, the resulting condition, if consistent with the M2-brane projection, would lead to a $1/4$ supersymmetric configuration that does not correspond to the M2-brane.

This is not the end of the story. For Chern-Simons theories one cannot rely on the naive analysis performed using the M-algebra. We know that the canonical structure of these theories is intricate. We should first check whether we are working in a degenerate sector or in a generic one, and use Dirac's formalism thoroughly to determine the exact form of the supersymmetry algebra on the constraint surface. This was partially done for 5D CS supergravity in \cite{MTZ}. A necessary step to put our conclusions on a firm ground involves a generalization of this analysis to the 11D case, which is not an easy job. It is still intriguing to figure out how the inconsistency between the $\Gamma$ matrix structure of the algebra and the supersymmetry transformation laws shall be solved. For this to happen, the actual $\Gamma$ matrix structure should change after the Dirac analysis. This would lead to the very interesting scenario in which the starting point might be quite a rather strange looking CS theory whose constrained algebra looks like the M-algebra. This seems very hard to implement.

In what follows, we present an alternative route to couple a CS theory to a brane, taking as a model the electromagnetic coupling to a the worldline of a point-charge ($0$-brane).  Starting from the observation that the electromagnetic coupling is also the integral of a CS form, the coupling is generalized to higher-dimensional branes and to non-Abelian connections. The resulting structure may not be the most general form, but it has the advantage that it exploits the geometric features of the CS forms to bring about the interactions (no metric required, topological origin, quantized charges, etc.).

\section{CS  actions as brane coupling}

A Chern-Simons action is a functional for a Lie-algebra valued one-form $\mathcal{A}$, defined in a topological space of dimension $D=2n+1$,
\begin{equation} \label{CS-2n+1}
I_{2n+1}[\mathcal{A}] = \frac{\kappa }{n+1}\int_{\Gamma^{2n+1}}\, \sum_{k=0}^n\, c_k\, \langle  \mathcal{A}^{2k+1} \wedge (d\mathcal{A})^{n-k} \rangle\;,
\end{equation} 
where $\langle \cdots \rangle$ denotes the symmetrized trace in some representation of the Lie algebra, $\mathcal{A}^p$ means $\mathcal{A} \wedge \cdots \wedge \mathcal{A}$ ($p$ times), $(d\mathcal{A})^q$ should be analogously understood as $d\mathcal{A} \wedge \cdots \wedge d\mathcal{A}$ ($q$ times), $c_k$ are specific coefficients ($c_0 = 1$), and $\kappa$ is a constant, known as the level of the theory. The fundamental difference between CS theories and the vast majority of physical actions is the absence of a metric structure and of dimensionful parameters in the former. This makes the theory simultaneously scale invariant, covariant under general coordinate transformations, and background independent.

The simplest example of a CS action is the familiar minimal coupling between an electric point charge and the electromagnetic potential, 
\begin{eqnarray} \label{jA}
I_{Int} = \int_{M^D} j^{\mu} \mathcal{A}_{\mu} d^Dx \;. 
\end{eqnarray} 
Since the current density $j^{\mu}$ has support on the worldline of the charge, (\ref{jA}) can also  be written as an integral over a $(0+1)$-dimensional manifold,which corresponds to the case $n=0$ in Eq. (\ref{CS-2n+1}),
\begin{eqnarray} \label{CS-0+1}
I_{0+1}[\mathcal{A}] = \kappa \int_{\Gamma^1} \langle \mathcal{A} \rangle \;.
\end{eqnarray} 
Here the manifold $\Gamma^1$ is the worldline of the charge, a 1-dimensional  submanifold embedded in the higher-dimensional space $M^D$, which is identified as the spacetime. The $D$-dimensional embedding spacetime may have a metric which induces a natural metric on the worldline, but this metric is not necessary to construct the action.  

In \cite{Teitelboim:1985ya}, Bunster analyzed the generalization of (\ref{jA}) to describe the coupling between a $(p-1)$-brane to a gauge potential, with an interacion of the form
\begin{eqnarray} \label{jA-p}
I_{Int} = \int_{M^D} j^{\mu_1 \mu_2 \cdots \mu_p}\, A_{\mu_1 \mu_2 \cdots \mu_p} d^Dx \;.
\end{eqnarray} 
He showed that this form of minimal coupling can only be defined (for any $p > 1$) if the connection is Abelian, {\it i.e.}, it transforms as $A \to A+d\Lambda $, where $\Lambda$ is a (real valued) $(p-1)$-form. The extension to non-Abelian connections was shown to be inconsistent due to the noncommutativity of the Hamiltonian at different times. As we show below, this obstruction does not arise if the branes couple to CS forms, Abelian or otherwise.

\subsection{0+1 CS theories}

As emphasized in \cite{z-BsAs}, the same expression (\ref{jA}) can also be interpreted as the action, in Hamiltonian form, for an arbitrary mechanical system of finitely many degrees of freedom \cite{DJT}. Therefore all mechanical systems are also examples of CS theories. Moreover, the Bohr-Sommerfeld rules of quantum mechanics, as well as Dirac's quantization rule for electric-magnetic charges, can be seen as consequences of the topological origin of CS theories, the Chern classes. So, CS theories are far from exceptional, they seem to be rather commonplace in physics.

For most Lie groups of physical interest (unitary, orthonormal), the generators are traceless and therefore $\langle \mathcal{A} \rangle=0$. The only important exception is the $U(1)$ group, and therefore one should look at the coupling between an electric charge $e=\kappa$  and the electromagnetic field,
\begin{equation}  \label{eA}
I= e \int_{\Gamma^1} \mathcal{A}_{\mu}(z)\, dz^{\mu} \;,
\end{equation}
where $z^{\mu}$ are the embedding coordinates of the worldline $\Gamma^1$, giving the position of the charge in $M^D$.  The interesting point is that the electromagnetic interaction is a model that captures the essential features of the coupling between higher-dimensional CS theories and branes.

The point charge is described by a delta function with support at the position of the charge on the spatial section $x^0=$ constant. The interaction term is
\begin{equation}  \label{jA'}
I= \int_{M^D} j^{(D-1)}_0\wedge \mathcal{A} \;,
\end{equation}
where  
\begin{equation}
j^{(D-1)}_0=\kappa \delta^{(D-1)}(x-z) d\Omega^{D-1} \;.
\end{equation} 
Here $d\Omega^{D-1}$ is the volume form of the spatial section in the rest frame of the charge.  
The action (\ref{eA}) by itself can be varied with respect to the embedding coordinates $z^{\mu}$ which, in the mechanical interpretation, are the enlarged phase space coordinates, $z^\mu \leftrightarrow (p^i, q_i, t)$ \cite{z-BsAs}. This means, in particular, that the embedding space must be odd-dimensional. This underscores the fact that a CS theory in a spacetime of dimension $D=(2n+1)$ can be naturally coupled to a $0+1$ Chern-Simons action defined on a one-dimensional worldline. This idea may be easily generalized to include CS actions for all lower (odd-) dimensional worldvolumes generated by $2p$-branes, with $p<n$, as we show next.

\subsection{(2n+1)-dimensional Abelian CS theories and 2p branes}

Comparing (\ref{CS-2n+1}) with the expression for the coupling between a point charge and the electromagnetic potential (\ref{CS-0+1}), it is clear that a $(2p+1)$ CS action, can be viewed as the coupling between the connection and a $2p$-brane \cite{Mora-Nishino,z-BsAs}. One is then led to consider the general coupling between an Abelian connection $\mathcal{A}$ and external sources with support on the $(2p+1)$-worldvolumes of all possible $2p$-branes that can be embedded in $M^{2n+1}$,
\begin{equation}
I_{2n+1}[\mathcal{A}] \; = \; \int_{\Gamma^{2n+1}} \sum_{p=0}^n j_{2p}^{(2n-2p)} \wedge {\cal C}_{2p+1} \; = \; \sum_{p=0}^n \kappa_p \int_{\Gamma^{2p+1}} {\cal C}_{2p+1} \;.
\label{generalcoupling}
\end{equation} 
Here the levels $\kappa_p$ are independent dimensionless coupling constants that can be identified with the ``electromagnetic'' charges.\footnote{Local gauge invariance of the Chern-Simons form guarantees that $I_{2n+1}[\mathcal{A}]$ is gauge invariant provided the currents $j_{2p}^{(2n-2p)}$ are closed (conserved), $dj_{2p}^{(2n-2p)} = 0$. Under quite general arguments, analogous to Dirac's for the quantization of the electric/magnetic charges, it can be shown that these charges must also be quantized.} For simplicity we set $\kappa_n=1$ here. The simplest rendering of this form is $p=0$, $n=1$: a point charge acting as the source of a $2+1$ Abelian CS connection. The action reads
\begin{equation}
I[\mathcal{A}]=  \int_{\Gamma^{2+1}}  \left[\frac{1}{2} \mathcal{A}\wedge d\mathcal{A} +  j_0^{(2)} \wedge \mathcal{A}\right] \;.
\end{equation}
Assuming $\Gamma^{2+1}$ to be compact and without boundary, the action can be varied with respect to $\mathcal{A}$. The field equation reads, not surprisingly,
\begin{equation}
\mathcal{F}=j_0^{(2)} \;,
\end{equation}
where $j_0^{(2)}$ is the 2-form charge density describing a point charge at rest,
\begin{equation}
j_0^{(2)}=\kappa_0 \delta^{(2)}(\vec{z})dx \wedge dy \;.
\end{equation} 
The field $\mathcal{A}$ is given by
\begin{equation}
\mathcal{A}=\frac{\kappa_0}{2\pi}d\phi \;, 
\end{equation} 
as shown by direct integration of the field equation $d\mathcal{A}=\kappa_0 \delta(x,y)dx \wedge dy$ on a disc, and using Stokes' theorem on a manifold that is topologically $\mathbb{R}^2-\{0\}$.  This source produces a magnetic field ($\mathcal{F}_{0i}=0$) concentrated along the worldline of the charge, like an infinitely thin solenoid (with the only pecularity that the solenoid is infinitely long in the time direction) \cite{JT-S}. So, this configuration is the electromagnetic field produced by a magnetic point source (monopole).
\vskip1mm

Similarly, a $(2n+1)$-CS form couples to the worldvolume of a charged $2p$-brane through the interaction
\begin{equation}\label{CScoupling}
I[\mathcal{A}]=  \int_{\Gamma^{2n+1}} j_{2p}^{(2n-2p)}\wedge {\cal C}_{2p+1} \;, 
\end{equation} 
where $j_{2p}^{(2n-2p)}$ stands for a $(2n-2p)$ form with support on the worldvolume of the $2p$ brane embedded in the $(2n+1)$-dimensional spacetime, and the field equations read 
\begin{equation}
\mathcal{F}^n = \sum_p  j_{2p}^{(2n-2p)} \wedge \mathcal{F}^p \;.
\end{equation} 
where $\mathcal{F}^k = \mathcal{F} \wedge \cdots \wedge \mathcal{F}$ ($k$ times).

\subsection{Coupling of non-Abelian CS actions to $2p$-branes}

The above construction can be extended to non-Abelian connections simply allowing ${\cal C}_{2p+1}$ to be a $(2p+1)$ CS form for the same non-Abelian connection\footnote{Note that although $\mathcal{A}$ may be a non-Abelian connection, ${\cal C}_{2p+1}$ is in the center of the algebra and hence commuting. In this way, the obstruction presented in \cite{Teitelboim:1985ya} can be circumvented.} $\mathcal{A}$. However, the generalization meets an important new constraint: the invariant tensor $\tau_{a_1 a_2 \cdots a_{n+1}}$ of a given Lie algebra, with generators ${\bf G}_{a}$, $a=1, 2, \cdots r$,
\begin{equation}
\tau_{a_1 a_2 \cdots a_{n+1}}:= \langle {\bf G}_{a_1} {\bf G}_{a_2} \cdots {\bf G}_{a_{n+1}} \rangle \;,
\end{equation} 
required for the CS action in $2n+1$ dimensions (see, for example, eq.(\ref{invT})), may not be defined for all values of $n$. It is an open problem how many invariant tensors of a given rank there exist for a given Lie algebra. This puts a severe restriction on the kinds of allowed couplings between a non-Abelian connection $\mathcal{A}$ and a $2p$-brane. Generically, one could write (\ref{CScoupling}) as in the previous case, but since there is no guarantee that a given Lie algebra admits an invariant tensor of a certain rank, many CS forms ${\cal C}_{2p+1}$ may vanish identically. 
\vskip1mm

An alternative possibility is that the $2p$-brane couples to a $(2p+1)$-CS form constructed with the invariant tensor for a \textit{subalgebra} of the Lie algebra defining the local symmetries of the theory in the embedding space $M^{2n+1}$. In fact, the presence of the brane generically produces a topological defect that partially breaks the original gauge symmetry. The surviving symmetry forms a subalgebra that admits an invariant tensor that can be used to construct a CS form on the worldvolume of the brane/defect. This was observed to occur in the presence of a codimension $2$ topological defect \cite{AWZ}. There, the  defect breaks the gauge symmetry $SO(D-1,2)$ down to $SO(D-2,1)$, giving rise to a gravitational action in $D-1$ dimensions out of a topological invariant in $D+1$ dimensions.
\vskip1mm

Another interesting feature of this mechanism of symmetry breaking by a $2p$-brane is this: suppose one couples a connection for the AdS algebra in $2n+1$ dimensions; the worldvolume of the brane\footnote{These arguments may be extended to the case of spacelike worldvolumes. The fate of supersymmetry is unclear in this case, though.}, is a manifold of dimension $2p+1$, and the maximal symmetry of the tangent space is $SO(2p,2)$. Since the number of components of a spinor representation goes as $2^{[D/2]}$, for every reduction by two in the dimension of the brane, there is a reduction by half in the number of components of the possible Killing spinors admitted by the configuration. This means that one can expect to generate $1/2$, $1/4$ (in general $1/2^k$) BPS states in this manner.
\vskip1mm

One alternative to the breaking would occur if the fermions in the space with a defect are combined in complex representations. For instance, starting from 11D and an $osp(32|1)$ real spinor with $32$ components, the presence of a codimension two defect would break the spacetime symmetry down to $so(8,2) \times so(2)$ admitting a complex spinor with $16$ components. The supersymmetric extension of the AdS group in 9D is $SU(8,8|1)$. Topological defects will host Killing spinors living in representations of the latter group and this generically implies the breaking of a fraction of the original supersymmetries. For particular values of the parameters, however, it might happen that the defect preserves all the supercharges (see the example below).

An interesting case that deserves further discussion is that of a membrane coupled to an 11D CS theory for the M-algebra. It is not hard to be tempted to identify such an object as the M2-brane. Notice that its coupling to the gauge connection is given through a 3D CS action based at most on the maximal supersymmetric extension of $so(2,1) \times so(8)$. This term is reminiscent of the action for multiple M2-branes recently unveiled by Bagger and Lambert \cite{BL}. No doubt that there are important differences, such as the presence of extra scalar fields (and their supersymmetric partners) and the fact that the CS Lagrangian in the BL theory is based upon a 3-algebra. Our approach attempts to address how these branes couple to the 11D fields while BL theory aims at describing the dynamics of multiple M2-branes on their own. They are not on equal footing. However, we find striking that our independent proposal for the introduction of M2-branes in a CS theory based on the M-algebra possesses these similarities and consider that this is a worth exploring avenue for further research.

\subsubsection{Example: Topological defects}

Branes are, in a broad sense, topological obstructions, like boundaries and defects. They restrict the continuous differentiable propagation on the manifold, which have topological consequences for the allowed orbits and for the spectrum of the differential operators. CS systems are particularly sensitive to the topological structure of the spacetime on which they are defined, and therefore the coupling between a connection dynamically governed by a CS action and a brane is necessarily nontrivial. Instead of developing a general theory for this problem, we illustrate this with an example. The discussion will remain at an introductory level and the reader is encouraged to look for the relevant sources as they become available.

Consider a $(2n+1)$-dimensional AdS spacetime where a point has been removed from the spatial section. The evolution in time of the missing point is a removed 1-dimensional worldline. The resulting topology is not that of AdS and it allows for nontrivial winding numbers for $S^{D-2}$ spheres mapped onto the spacetime. In principle, there could be an angular defect concentrated on the removed line, which measures the strength of the singularity. The defect is produced by an identification in the angular directions whereby the solid angle $\Omega_{D-2}$ of the $S^{D-2}$ sphere is shrunk to $(1-\alpha)\Omega_{D-2}$ The metric produced by this angular defect can be written as
\begin{equation}
ds^2=-(1+\rho^2)dT^2 + (1+\rho^2)^{-1}d\rho^2 + \alpha^2 \rho^2 d\Omega_{D-2}^2 \;.
\end{equation} 
It is straightforward to check that this metric has a naked curvature singularity in the angular components of the Riemann tensor,
\begin{equation}\label{Dim-cont-BH}
R^{\alpha \beta}_{\;\;\;\;\; \theta \phi}= \left[ -1 + \frac{\sqrt{1+M}}{r^2} \right] \delta^{\alpha \beta}_{\;\;\;\;\; \theta \phi} \;,
\end{equation} 
while the remaining components are those of a constant curvature, $R^{0r}_{\;\;\;\;\; 0r}=-1$, $R^{0\alpha}_{\;\;\;\;\; 0\beta}=R^{r\alpha}_{\;\;\;\;\; r\beta}=-\delta^{\alpha}_{\;\;\; \beta}$. This looks like the standard $(2n+1)$ AdS CS black hole (in $r(\rho)$ radial coordinate) \cite{Banados:1993ur}. However, for $0 < \alpha < 1$, the (dimensionless) mass parameter corresponds to a ``negative mass black hole", $-1 < M(\alpha) < 0$, which is just a naked singularity \cite{Olivera}. 

The resulting space does not admit Killing spinors and the singularity cannot correspond to a BPS configuration, except for $D=3$ in the limit when the angular defect becomes maximal ($\alpha \rightarrow 1$). This special case, $M=0$, corresponds to a massless 2+1 black hole and the space now admits half of the Killing spinors of the AdS spacetime. In this case, the defect can legitimately be called a BPS 0-brane\footnote{For $D>3$ the massless black holes also have a curvature singularity and are not BPS, as shown in \cite{Aros:2002rk}.} \cite{Olivera}. 

The massless 2+1 black hole can also be generated through a particular identification by a Killing vector in AdS$_3$ space and this mechanism can be repeated in higher dimensions: starting with AdS$_{2n+1}$,  an identification with a Killing vector that has a fixed point generates a topological defect at the fixed point and breaks the symmetry from $SO(2n,2)$ down to $SO(2n-1,1)\times R$. The AdS space has maximal supersymmetry with $2^n$ component spinors; the topological defect can have at most $2^{n-1}$ local supersymmetries, that is $1/2$ BPS. Additional breakings generated by further identifications with Killing spinors, would reduce the supersymmetry to $1/4$, $1/8$, etc. \cite{Olivera,Jose}. This topological symmetry breaking was recently exploited in \cite{AWZ} to generate an effective Einstein-Hilbert action in four dimensions from a topological defect in a six-dimensional topological field theory. 

\section{Summary and gambling on future directions}

We have presented a proposal for the coupling of sources in CS theories. When the spacetime dimensions is $D \leq 3$, this amounts to the standard minimal gauge coupling. However, for higher dimensional CS theories this produces new interaction terms with a number of consequences on which this article just offers a first glance. The coupling is entirely given in terms of the connection of the original theory and does not require (the otherwise problematic) non-Abelian $p$-forms to couple directly with extended objects such as branes. We argue that this suggests the need to revisit the exploration of the BPS spectrum of CS theories, a certainly difficult subject, in a way that possibly circumvents naive obstructions to the existence of expected objects such as the M2-brane in a CS theory based on the M-algebra.

We have explored the existence of preons in CS theories. Following the algebraic reasoning introduced in \cite{BdAIPV}, we have shown that it applies to CS theories based on extensions of the AdS algebra, but needs some improvement for the case of Poincar\'e-based theories such as the CS supergravity for the M-algebra. We should emphasize that even if the integrability conditions of the local Killing spinor conditions are consistent with the equations of motion, it is still necessary to scrutinize the actual integrability of the equations, as there might be, for example, topological obstructions.

Our proposal for the introduction of sources implies a novel mechanism of symmetry breaking through the presence of defects in CS theories. Since the couplings are given in terms of CS forms with support in lower dimensional submanifolds, they will be written generally in terms of subalgebras of the original algebra.

The quantization of CS theories is an important problem. The case of $0+1$ is just the old Bohr-Sommerfeld quantization \cite{z-BsAs}. The $2+1$ case is well-understood: the path integral is given in terms of knot invariants \cite{Witten-BarNatan}. The corresponding quantization for higher dimensions remains an open problem.

The emergence of a dimensionful physical scale in the theory may arise through condensates of the form $\langle \mathcal{A} \bullet \mathcal{A} \rangle \neq 0$, where the bracket is an invariant tensor of the reduced (physical) symmetry. For instance, in a CS theory based on the AdS group broken down to the Lorentz group, $\langle \mathcal{A} \bullet \mathcal{A} \rangle$ reduces to $e^a_\mu\, e^b_\nu\, \eta_{ab} = \ell^2\, g_{\mu\nu}$. The remnant Lorentz symmetry suggests the possibility that the induced dynamics may be governed by the Einstein-Hilbert action. This is indeed the case in the scheme studied in \cite{AWZ}, and might be a promising avenue to explore the (still open) connection between CS theories and ordinary supergravity. 

\begin{acknowledgement}
We are pleased to thank Mokhtar Hassa\"\i ne, Olivera Mi\v{s}kovi\'{c} and Steve Willison, for many interesting discussions and comments. We are also thankful to Claudio Bunster for so many years of collaboration and friendship. This work is supported in part by MEC and FEDER (grant FPA2005-00188), by Xunta de Galicia (Conseller\'\i a de Educaci\'on and grant PGIDIT06PXIB206185PR), by the European Commission (grant MRTN-CT-2004-005104), and by the Spanish Consolider-Ingenio 2010 Programme CPAN (CSD2007-00042). JDE is a {\it Ram\'on y Cajal} Fellow. The Centro de Estudios Cient\'\i ficos (CECS) is funded by the Chilean Government through the Millennium Science Initiative and the Centers of Excellence Base Financing Program of Conicyt. CECS is also supported by a group of private companies which at present includes Antofagasta Minerals, Arauco, Empresas CMPC, Indura, Naviera Ultragas and Telef\'onica del Sur. This work was partially supported by Fondecyt grants $\# 1061291$ and $\# 7061291$.
\end{acknowledgement}


\begin{thebibliography}{99}

\bibitem{HTZ}
Hassa\"{\i}ne, M., Troncoso R., Zanelli, J.:
Eleven-dimensional supergravity as a gauge theory for the M-algebra.
Phys.\ Lett.\  B {\bf 596}, 132 (2004)

\bibitem{BdAIL}
Bandos, I.A., de Azc\'arraga, J.A., Izquierdo, J.M., Lukierski, J.:
BPS states in M-theory and twistorial constituents.
Phys.\ Rev.\ Lett.\  {\bf 86}, 4451 (2001)

\bibitem{BdAIPV}
Bandos, I.A., de Azc\'arraga, J.A., Izquierdo, J.M., Picon, M., Varela, O.:
BPS preons, generalized holonomies and 11D supergravities.
Phys.\ Rev.\  D {\bf 69}, 105010 (2004)

\bibitem{GGPR}
Gran, U., Gutowski, J., Papadopoulos, G., Roest, D.:
N=31, D=11.
JHEP {\bf 0702}, 043 (2007)

\bibitem{CT82}
Teitelboim, C.:
Quantum mechanics of the gravitational field.
Phys.\ Rev.\ D {\bf 25}, 3159 (1982)

\bibitem{Teitelboim:1985ya}
Teitelboim, C.:
Gauge Invariance For Extended Objects.
Phys.\ Lett.\  B {\bf 167}, 63 (1986) \\
Henneaux, M., Teitelboim, C.:
p-Form Electrodynamics.
Found.\ Phys.\ {\bf 16}, 593 (1986)

\bibitem{LecturesZ}
Zanelli, J.:
Lecture notes on Chern-Simons (super-)gravities. Second edition (February 2008).
arXiv:hep-th/0502193.

\bibitem{EZsort}
Edelstein, J.D., Zanelli, J.:
(Super-)Gravities of a different sort.
J.\ Phys.\ Conf.\ Ser.\  {\bf 33}, 83 (2006)

\bibitem{MTheory}
Hull, C.M., Townsend, P.K.:
Unity of superstring dualities.
Nucl.\ Phys.\  B {\bf 438}, 109 (1995) \\
Witten, E.:
String theory dynamics in various dimensions.
Nucl.\ Phys.\  B {\bf 443}, 85 (1995)

\bibitem{Troncoso:1997va}
Troncoso, R., Zanelli, J.:
New gauge supergravity in seven and eleven dimensions.
Phys.\ Rev.\  D {\bf 58}, 101703 (1998)

\bibitem{WittenMthParity}
Horava, P., Witten, E.:
Heterotic and type I string dynamics from eleven dimensions.
Nucl.\ Phys.\  B {\bf 460}, 506 (1996) \\
Witten, E.:
On flux quantization in M-theory and the effective action.
J.\ Geom.\ Phys.\  {\bf 22}, 1 (1997)

\bibitem{Horava}
Horava, P.:
M-theory as a holographic field theory.
Phys.\ Rev.\  D {\bf 59}, 046004 (1999)

\bibitem{PKT}
Townsend, P.K.:
P-brane democracy.
arXiv:hep-th/9507048.

\bibitem{HS}
Hatsuda, M., Sakaguchi, M.:
Wess-Zumino term for the AdS superstring and generalized Inonu-Wigner contraction.
Prog.\ Theor.\ Phys.\  {\bf 109}, 853 (2003)

\bibitem{dAIPV}
de Azc\'arraga, J.A., Izquierdo, J.M., Pic\'on, M., Varela, O.:
Generating Lie and gauge free differential (super)algebras by expanding
Maurer-Cartan forms and Chern-Simons supergravity.
Nucl.\ Phys.\  B {\bf 662}, 185 (2003) \\
de Azc\'arraga, J.A., Izquierdo, J.M., Pic\'on, M., Varela, O.:
Expansions of algebras and superalgebras and some applications.
Int.\ J.\ Theor.\ Phys.\  {\bf 46}, 2738 (2007)

\bibitem{EHTZ}
Edelstein, J.D., Hassa\"{\i}ne, M., Troncoso R., Zanelli, J.:
Lie-algebra expansions, Chern-Simons theories and the Einstein-Hilbert
Lagrangian. Phys.\ Lett.\  B {\bf 640}, 278 (2006)

\bibitem{MokRom}
Hassa\"{\i}ne, M., Romo, M.:
Local supersymmetric extensions of the Poincare and AdS invariant gravity.
JHEP {\bf 0806}, 018 (2008)

\bibitem{WO}
Witten, E., Olive, D.I.:
Supersymmetry Algebras That Include Topological Charges.
Phys.\ Lett.\  B {\bf 78}, 97 (1978)

\bibitem{MaxB}
Ba\~nados, M.:
The linear spectrum of OSp(32|1) Chern-Simons supergravity in eleven dimensions.
Phys.\ Rev.\ Lett.\  {\bf 88}, 031301 (2002)

\bibitem{MTZ}
Mi\v{s}kovi\'{c}, O., Troncoso, R., Zanelli, J.:
Canonical sectors of 5d Chern-Simons theories.
Phys.\ Lett.\ B {\bf 615}, 277 (2005)

\bibitem{z-BsAs}
Zanelli, J.:
Uses of Chern-Simons Actions.
In: Edelstein, J.D., Grandi, N., N\'u\~nez, C., Schvellinger, M. (eds.) 
Ten years of AdS/CFT, pp.115--129. American Institute of Physics, New York (2008)

\bibitem{DJT}
Dunne, G.V., Jackiw, R., Trugenberger, C.A.:
Topological (Chern-Simons) Quantum Mechanics.
Phys.\ Rev.\  D {\bf 41}, 661 (1990)

\bibitem{Mora-Nishino}
Mora, P., Nishino, H.:
Fundamental extended objects for Chern-Simons supergravity.
Phys.\ Lett.\  B {\bf 482}, 222 (2000) \\
Mora, P.: 
Chern-Simons supersymmetric branes.
Nucl.\ Phys.\ B {\bf 594}, 229 (2001)

\bibitem{JT-S}
Jackiw R., Templeton, S.:
How Superrenormalizable Interactions Cure Their Infrared Divergences.
Phys.\ Rev.\  D {\bf 23}, 2291 (1981) \\
Schonfeld, J.F.:
A Mass Term For Three-Dimensional Gauge Fields.
Nucl.\ Phys.\  B {\bf 185}, 157 (1981)

\bibitem{AWZ}
Anabal\'{o}n, A., Willison, S., Zanelli, J.:
General relativity from a gauged WZW term.
Phys.\ Rev.\  D {\bf 75} 024009 (2007) \\
Anabal\'{o}n, A., Willison, S., Zanelli, J.:
The Universe as a topological defect.
Phys.\ Rev.\  D {\bf 77} 044019 (2008)

\bibitem{BL}
Bagger J., Lambert, N.:
Gauge Symmetry and Supersymmetry of Multiple M2-Branes.
Phys.\ Rev.\  D {\bf 77}, 065008 (2008)

\bibitem{Banados:1993ur}
Ba\~nados, M.,Teitelboim, C., Zanelli, J.:
Dimensionally continued black holes.
Phys.\ Rev.\  D {\bf 49} 975 (1994)

\bibitem{Olivera}
Mi\v{s}kovi\'{c}, O., Zanelli, J.:
Coupling between AdS gravities and 2p-branes.
To appear (2008)

\bibitem{Aros:2002rk}
Aros, R., Mart\'{\i}nez, C., Troncoso, R., Zanelli, J.:
Supersymmetry of gravitational ground states.
JHEP {\bf 0205}, 020 (2002)

\bibitem{Jose}
Edelstein, J.D., Zanelli, J.:
In progress (2008)

\bibitem{Witten-BarNatan}
Witten, E.:
Quantum field theory and the Jones polynomial.
Commun.\ Math.\ Phys.\  {\bf 121}, 351 (1989) \\
Bar-Natan, D., Witten, E.:
Perturbative expansion of Chern-Simons theory with noncompact gauge group.
Commun.\ Math.\ Phys.\  {\bf 141}, 423 (1991)

\end{thebibliography}
\end{document}